# Reviewing National Cybersecurity Awareness for Users and Executives in Africa

Maria Bada

Cambridge Cybercrime Centre
Computer Laboratory,
Cambridge University UK
Global Cyber Security Capacity Centre,
University of Oxford, UK
Academy for Computer Science
and Software Engineering
University of Johannesburg
South Africa
Email: `maria.bada@cl.cam.ac.uk`

Basie Von Solms

Academy for Computer Science
and Software Engineering
University of Johannesburg,
South Africa
Email: `basievs@uj.ac.za`

Ioannis Agrafiotis

Department of Computer Science
University of Oxford
Email: `ioannis.agraotis@cs.ox.ac.uk`

*Abstract*—There is an unprecedented increase in cybercrime globally observed over the last years. One of the regions driving this increase is Africa, where significant financial losses are reported. Yet, citizens of African countries are not aware of the risks present in cyberspace. The design and implementation of national awareness campaigns by African countries to address this problem are in their infancy state, mainly due to the absence of capacity building efforts. As part of the Global Cybersecurity Capacity Centre (GCSCC) programme, we conducted a series of focus groups in six African countries, in order to assess their cybersecurity posture, a critical component of which is user and executive awareness of cyber risks. This paper is an extended version of previous work where an initial analysis of awareness for cyber risk in African countries was presented. In this extended version, we reflect on best practice approaches for developing national awareness campaigns and use these as a framework to analyse qualitative data from the focus groups. We discuss the current state of African countries with regards to the implementation of national cybersecurity awareness campaigns for users and executives, the main obstacles in combating cybercrime and conclude with recommendations on how African countries can identify and prioritise activities to increase their capacity regarding cybersecurity awareness.

## I. INTRODUCTION

Over the last years, there has been an unprecedented increase in cybercrime globally [1], [2], [3]. Africa is a region with one of the highest rates of cybercrime affecting the strategic, economic and social growth development of the region [4]. Reports suggest that, inter alia, estimated costs have soared up to $550 million for Nigeria, $175 million for Kenya and $85 for Tanzania [4].

Additionally, the growth in Internet use has been facilitated by high proliferation and adoption of mobile communications. Speedy diffusion and adoption have exposed the public to unprecedented individual security threats via the mobile platform [5].

Studies have confirmed that mobile phones have been used as a platform for distributing viruses as well as a transmission of viruses over Bluetooth services [6]. In some instances, mobile phones have been used to propagate hate speech as evidenced in Kenya after the December 2007 elections that fuelled ethnic violence [7].

One of the factors creating a permissive environment for cybercrime is the lack of awareness in the African public regarding risks when using cyberspace [4]. Additionally, the level of development of digital infrastructure in African countries directly influences their security posture. Reports suggest that cyber criminals rely on the very poor security habits of the general population [8] and urge policy makers to engage in awareness campaigns [4] since there is strong evidence that such initiatives can efficiently lower the success rate of cybercrime [9]. More specifically, there are white papers estimating that an investment in security awareness and training can potentially change user's behavior and reduce cyber-related risks by 45% to 70% [9].

It is evident that Cybersecurity Awareness is a very important step in the fight against cybercrime in Africa. For that reason, it is essential for any African country that intends to implement interventions in this area to have a holistic understanding of the level of Cybersecurity Awareness in that country. Towards this direction, there have been efforts to capture the status of Cybersecurity Awareness (understanding on cyber threats and risk, cyber hygiene, and appropriate response options) in Africa [10], and in general, the findings suggest that the absence of awareness campaigns regarding cybersecurity and Internet safety create a lax environment for information security [10]. In 2016, only 11 (20.3%) out of 54 countries had implemented cybersecurity (CS) laws and regulations [11]. Additionally, the lack of an adequately skilled workforce on cybersecurity can impose great challenges to many African countries.

This paper is an extended version of previous work [1]

where an initial analysis of awareness for cyber risk in African countries was presented. In this extended version, we analyse qualitative data from six African countries that was collected when applying the Cybersecurity Capacity Maturity Model for Nations (CMM) developed by the Global Cybersecurity Capacity Centre (GCSCC) at the University of Oxford [12]. We reflect on best practice approaches for developing campaigns and draw conclusions on what the current state of African countries is regarding awareness in risks from cybercrime, what are the main obstacles in combating cybercrime and what actions countries should prioritise in order to increase awareness of risks from cybercrime in their population.

In what follows, Section II provides a literature and best practice review on developing cybersecurity awareness campaigns and existing efforts in Africa. Section III provides a brief overview of the CMM and the CMM methodology when deployed in a country. Section IV describes the results from the CMM reviews in six African countries and our analysis of the qualitative data obtained from focus groups during these reviews. As this paper concentrates on Cybersecurity Awareness, which is one component of the CMM, only the results of this component will be discussed. No countries will be referenced, but a general overview of the outcome will be described. Section V discusses the results of our analysis and Section VI concludes the paper.

II. CYBERSECURITY AWARENESS RAISING CAMPAIGNS

According to the UK Her Majesty's Government (HMG) Security Policy Framework [13], it is government's role to raise cybersecurity awareness within a country. "*People and behaviours are fundamental to good security. The right security culture, proper expectations and effective training are essential. Everyday actions and the management of people, at all levels in the organisation, contribute to good security. A strong security culture with clear personal accountability and a mature understanding of managing risk, responsibility and reputation will allow the business to function most effectively*".

Awareness presentations are "*intended to allow individuals to recognize IT security concerns and respond accordingly. In awareness activities, the learner is the recipient of information, whereas the learner in a training environment has a more active role. Awareness relies on reaching broad audiences with attractive packaging techniques. Training is more formal, having a goal of building knowledge and skills to facilitate the job performance*" [14].

Awareness is used to stimulate, motivate, and remind the audience what is expected of them [15]. This is an important aspect of cybersecurity policy or strategy because it enhances the knowledge of users about security, changes their attitude towards cybersecurity, and their behaviour patterns.

A. Developing cybersecurity awareness raising campaigns

There is an abundance of best practice approaches describing principles in designing and implementing an awareness-raising campaign. Little emphasis, however, was put on how to strategically decide the areas where awareness campaigns should focus. National Institute of Standards and Technology (NIST) [16] is one of the pioneers in this field. Their framework provides three alternatives on how organisations should be structured, detailing for each category the processes for an effective and efficient campaign.

For all three approaches, namely centralised, partially decentralised and fully decentralized, NIST provides information on how a 'needs assessment' should be conducted; a strategy should be developed; an awareness training program be designed; and an awareness program be implemented. The key criteria to decide which approach an organisation should adopt are the size of the organisation, similarities in missions between different departments, knowledge of the topics into question and how spread the geographical area where campaigns will be implemented is.

Focusing on the design and implementation of awareness-raising campaigns, literature suggests that successful awareness campaigns need to be a 'learning continuum' [15], commencing from awareness, evolving to training and resulting in education. According to Organisation of American States (OAS) [17], it is of paramount importance that stakeholders from the public and private sector, Non-profit Government Organisations (NGOs), and technology and finance corporations to be involved. Once stakeholders are identified, the next steps in the OAS model provide instructions on how to define the goals of the campaign, the audience it targets and the strategy via which the campaign will be implemented.

Even by following best practise, several difficulties exist when it comes to creating a successful campaign: a) not understanding what security awareness really is; b) a compliance awareness program does not necessarily equate to creating the desired behaviours; c) usually there is lack of engaging and appropriate materials; d) usually there is no illustration that awareness is a unique discipline; e) there is no assessment of the awareness programmes [18]; f) not arranging multiple training exercises but instead focusing on a specific topic or threat does not offer the overall training needed [19].

Perceived control and personal handling ability, the sense one has that he/she can drive specific behaviour, has also been found to affect the intention of behaviour but also the real behaviour [20]. Culture is another important factor for consideration when designing education and awareness messages [21] as it can have a positive security influence to the persuasion process. Moreover, even when people are willing to change their behaviour, the process of learning a new behaviour needs to be supported [21].

Messages and advertisements are usually preferred when they match the cultural theme of the message recipient. Overall, a campaign should use simple consistent rules of behaviour that people can follow. This way, their perception of control will lead to better acceptance of the suggested behaviour. Moreover, even when people are willing to change their behaviour, the process of learning a new behaviour needs to be supported [21].

B. Cybersecurity awareness campaigns in Africa

A review in cybersecurity policies in African countries [22] shows that awareness raising is key issue either as a separate factor or as part of the role of the proposed National CSIRT. A cybersecurity policy and strategy may not be in place yet for all countries in Africa. However, there are already a number of organisations that have identified the need for continental coordination and increased cybersecurity awareness including the African Information Society Initiative (UNECA/AISI) [23], The Internet Numbers Registry for Africa (AfriNIC) [24],

ITU/GCA [25], Interpol, The Southern African Development Community (SADC) [26] and ISG-Africa [27].

There are existing efforts in Africa such as the ISC Africa [28]. This is a coordinated, industry and community-wide effort to inform and educate Africa's citizens on safe and responsible use of computers and the Internet, so that the inherent risks can be minimised and consumer trust can be increased. Also, Parents' Corner Campaign [29] is intended to co-ordinate the work done by government, industry and civil society. Its objectives are to protect children, empower parents, educate children and create partnerships and collaboration amongst concerned stakeholders.

Recently Facebook has also announced partnerships with over 20 non-governmental organisations and official agencies from the DRC, Ghana, Kenya, Nigeria and South Africa in support of Safer Internet Day (SID) marked on 6 February [30]. SID advocates making the internet safer, particularly for the youth, and is organised by the joint Insafe-INHOPE network with the support of the European Commission and funded by the Connecting Europe Facility programme (CEF).

Usually, most of official awareness-campaign sites include advice, which usually comes from security experts and service providers, who monotonically repeat suggestions such as use strong passwords. Such advice pushes responsibility and workload for issues that should be addressed by the service providers and product vendors onto users. One of the main reasons why users do not behave optimally is that security systems and policies are often poorly designed [31]. There is a need to move from awareness to tangible behaviours.

*C. Cybersecurity awareness raising for executives*

The view that executives are often not sufficiently prepared to handle cybersecurity risk has raised concerns in boardrooms nationwide and globally. Even if companies increase their investments in security, we see more and more serious cyber attacks. The main concern is whether executives are prepared to make the right cybersecurity investment decisions and develop effective cybersecurity strategies [32].

Most executives realize the threat cyber risk represents to their organisation and to the nation's economy and are being found liable for cyber systems failures. Governing agencies are taking regulatory action against boards and management with the full support of the courts [33].

To improve the situation, companies need to address two issues. First, directors need to have basic training in cybersecurity that addresses the strategic nature, scope, and implications of cybersecurity risk. Second, top management needs to provide meaningful data about not just the state of data security as defined by viruses quarantined or the number of intrusions detected, but also about the resilience of the organisation's digital networks [32].

Developing a common language for management and corporate directors to discuss cybersecurity issues is also important. Digital security specialists, like all subject-area experts, must be able to communicate effectively with board members and other leaders. Information security executives must be capable to present information at a level and in a format that is accessible to non-technical corporate directors [32].

Both management and directors need to be aware of:

1) the limitations of security (no practical cybersecurity strategy can prevent all attacks) and
2) the need for resilience (strategies to sustain business during a cyberattack and to recover quickly in the aftermath of a breach).

This means that having strategies which will ensure sustainability of business during a cybersecurity breach and quick recovery in its aftermath, is important. Networks constantly change, so tracking cyber risks and vulnerabilities over time and adapting accordingly is essential [32].

Additionally, the involvement by business executives ensures that possible adverse impacts from security incidents are viewed from a bottom-line as well as from an asset valuation perspective [34]. In response to the gaps mentioned above, executives follow two different paths of cyber governance. First, they add a cybersecurity expert to the board and second, they assess the cybersecurity maturity of the organization or nation against accepted standards such as NIST [34].

In Africa, on the issue of who attends cybersecurity awareness training, the responses showed that although 70 percent of core business staff in organisations attended cybersecurity awareness training, and there is significant attendance by other groups of participants, such as supervisors/functional managers (59 percent) and middle management (57 percent) [35].

According to the ISACA State of cybersecurity report published in 2017 [36], globally just 21 percent of chief information security officers (CISO) report to the chief executive officer (CEO) or the board, while 63 percent report through the chief information officer (CIO). This latter reporting structure, which is even more common in Africa, positions security as a technical issue rather than a business concern, reducing the scope of action and effectiveness of any cybersecurity initiatives.

III. THE CYBERSECURITY CAPACITY MATURITY MODEL FOR NATIONS (CMM)

The CMM developed by the Global Cybersecurity Capacity Centre (GCSCC) [12] at the University of Oxford is a comprehensive framework which assesses the cybersecurity capacity maturity of capabilities which are fundamental to building resilience of a country over 5 different dimensions: 1) Cybersecurity Policy and Strategy; 2) Cyber Culture and Society; 3) Cybersecurity Education, Training and Skills; 4) Legal and Regulatory Frameworks; 5) Standards, Organisations, and Technologies.

Every Dimension consists of a number of Factors which describe what it means to possess cybersecurity capacity. Each Factor is composed of a number of Aspects that structure the Factor's content. Each Aspect is composed of a series of indicators within five stages of maturity. These indicators describe the steps and actions that must be taken to achieve or maintain a given stage of maturity in the aspect/factor/dimension hierarchy. These 5 maturity stages are: 1) Start up; 2) Formative; 3) Established; 4) Strategic; 5) Dynamic. The progressive nature of the model assumes that lower stages have been achieved before moving to the next.

The five stages are defined as follows:

1) Start-up: at this stage either no cybersecurity maturity exists, or it is very embryonic in nature. There might

be initial discussions about cybersecurity capacity building, but no concrete actions have been taken. There is an absence of observable evidence of cybersecurity capacity at this stage.
2) Formative: some aspects have begun to grow and be formulated, but may be ad-hoc, disorganised, poorly defined - or simply new. However, evidence of this aspect can be clearly demonstrated.
3) Established: the indicators of the aspect are in place, and functioning. However, there is not well thought-out consideration of the relative allocation of resources. Little trade-off decision-making has been made concerning the relative investment in this aspect. But the aspect is functional and defined.
4) Strategic: at this stage, choices have been made about which indicators of the aspect are important, and which are less important for the particular organisation or state. The strategic stage reflects the fact that these choices have been made, conditional upon the state's or organisation's particular circumstances.
5) Dynamic: at this stage, there are clear mechanisms in place to alter strategy depending on the prevailing circumstances such as the technological sophistication of the threat environment, global conflict or a significant change in one area of concern (e.g. cybercrime or privacy). Dynamic organisations have developed methods for changing strategies in-stride. Rapid decision-making, reallocation of resources, and constant attention to the changing environment are features of this stage.

The assignment of maturity stages is based upon the evidence collected, including the general or average view of accounts presented by stakeholders, desktop research conducted and the professional judgment of GCSCC research staff. Using the GCSCC methodology recommendations are provided as to the next steps that might be considered by a nation to improve cybersecurity capacity.

In this paper, we focus on the factor 'Cybersecurity Awareness Raising' (shown in detail in Figure 3 and Figure 4 in the Appendix section). The aspects, within this factor are 'Awareness Raising Programmes' and 'Executive Awareness Raising' with various indicator specialisations for every maturity stage. The aspect 'Awareness Raising Programmes' examines the existence of a national coordinated programme for cybersecurity awareness raising, covering a wide range of demographics and issues, while the aspect 'Executive Awareness Raising' examines efforts raising executives' awareness of cybersecurity issues in the public, private, academic and civil society sectors, as well as how cybersecurity risks might be addressed. The CMM model was developed by conducting systematic reviews on best practice approaches which are publicly available, as well as consulting experts from various disciplines.

According to the CMM, the aspect 'Awareness Raising Programmes' will be measured to be on a Start-up stage of maturity if the indicator 'The need for awareness of cybersecurity threats and vulnerabilities across all sectors is not recognised, or is only at initial stages of discussion' is met and indicators from the next level are absent. This stage of maturity is comprised only by this one indicator. In order to be at the formative stage of maturity, the next two relevant Indicators must be met. As seen in Appendix, the number of Indicators may differ between maturity stages. In order to elevate a country's cybersecurity capacity maturity, all of the indicators within a particular stage will need to be fulfilled.

The first version of the CMM was finalized in 2014 [37] and the revised edition was published in 2017 [12]. So far, the CMM has only been deployed on the national level (rather than at the company/enterprise level), and 54 countries have been fully evaluated through engagement and collaboration with the host country.

*A. The CMM implementation methodology*

The process by which a (host) country is assessed is as important as the model itself. This process actually forms the basis of the whole review/research methodology on which a country review is based. This process forms and motivates the underlying research and application methodology of a CMM review and provides scientific validity to the results coming from a review  the process guarantees the validity and verification of the outputs. The first step of a country review is to identify a country host  that is the body which is responsible for all logistical arrangements in the host country.

The CMM employs a focus group methodology since it has been acknowledged to offer a rich set of data compared to other qualitative approaches [38], [39], [40]. Like interviews, focus groups are an interactive methodology with the advantage that during the process of collecting data and information diverse viewpoints and conceptions can emerge. It is a fundamental part of the method that rather than posing questions to every interviewee, the researcher(s) should facilitate a discussion between the participants, encouraging them to adopt, defend or criticise different perspectives [41].

Stakeholders are identified based on their expertise in each one of the components of every Dimension of the CMM. There will be 9 stakeholder (cluster) groups planned for each CMM review (Table I). For example, for the specific factor Cybersecurity Awareness Raising, academia, civil society groups, internet governance experts, one or more representatives from Universities, and Internet societies will be invited to take part in the focus group. This group of representatives are called a stakeholder (cluster) group.

While these stakeholder clusters are useful as a guiding structure for conducting the assessments, selecting the participants in country to engage with the CMM would be challenging without a thorough understanding of the complex network of domestic actors. Therefore, in order to gain a more thorough understanding of the domestic context in which the model will be applied, we worked alongside either international organisations with knowledge of such domestic dynamics or directly with a host ministry or organisation within a country. Additionally, we complemented that process by conducting desk research on stakeholders that might provide valuable insight into domestic cybersecurity capacity and then make recommendations to the host team.

Focus group sessions are led by the CMM Review Team. The assessments are typically conducted over the course of three to four days for 1.5 to 2 hours for each cluster. The host team would indicate in the invitation which cluster the participant will engage with. The CMM review team will lead the discussion facilitation with the different participants in order to aid the selection of stages of maturity in each category, factor and dimension. Due to the depth and nuance of cybersecurity capacity within the CMM, it would be impossible to go through the entire model with each stakeholders cluster. Therefore, each cluster responds to two dimensions of the CMM, depending on their relevant expertise. The participants should be able to provide or indicate evidence supporting their selection, so that subjective responses are minimised. If a country does not fulfill all of the criteria within a particular stage of maturity, the previous stage is selected, while noting which particular elements of capacity are missing for achieving the proceeding stage. This nuance is key, as it allows for more flexibility in understanding existing capacity, rather than assigning a stage of maturity that does not account for subtle variations.

TABLE I. Stakeholder groups participating in focus groups

| Stakeholder (cluster) groups |
|---|
| Academia, Civil Society Groups, and Internet Governance Representatives |
| Criminal Justice and Law Enforcement |
| Critical National Infrastructure |
| CSIRT and IT Leaders from Government and the Private Sector |
| Defense and Intelligence Community |
| Government Ministries |
| Legislators/Policy owners |
| Private Sector and Business |
| Cyber Task Force |

Additionally, the CMM review team would be facilitating the discussion, trying to keep the discussion on track without influencing the opinions of the group, but also avoiding only a few of the participants dominating the discussion [42], [43].

The consultations result in a comprehensive report indicating the relevant Maturity stage for all Factors and Aspects in all Dimensions. A comprehensive set of Recommendations is also provided to indicate to the country how to improve capacity and progress onto the next stages of maturity.

## IV. CMM RESULTS FOR AWARENESS RAISING IN AFRICA

In Africa, a team from the GCSCC has reviewed and evaluated 6 countries based on the CMM and following the methodology described in Section III. These countries were selected for a review at the time because they were in the process of drafting a national cybersecurity strategy. Therefore, the review would assist this process. These reviews have been conducted during the period June 2015 to January 2018.

Regarding the aspect 'Awareness Raising Programs' and 'Executive Awareness Raising', 12 focus groups have been conducted in total. The stakeholders who participated in the focus groups are from the following sectors: Public Sector Entities; Legislators/Policy Makers; Criminal Justice and Law Enforcement; Armed Forces; Academia; Civil Society; Private Sector; CSIRT and IT Leaders from Government and the Private Sector; Critical national infrastructure; Telecommunications Companies; and Finance Sector. Each focus group session had approximately 10-15 stakeholders and lasted on average 2 hours.

In order for the stakeholders to provide evidence on how many indicators have been implemented by a nation and to determine the maturity level of every aspect of the model, a consensus method is used to drive the discussions within sessions. During focus groups, researchers use semi-structured questions to guide discussions around indicators. During these discussions, stakeholders should be able to provide or indicate evidence regarding the implementation of indicators, so that subjective responses are minimised.

### A. Analysis of maturity level data

Figure 1 illustrates the results from the six African countries of the CMM maturity levels for the cybersecurity awareness raising dimension. Three countries have been identified to be at a start-up stage of maturity, two countries have been identified at a formative stage and one at a start-up stage with few of the indicators from the formative stage of maturity being present (this is denoted as from start-up to formative in the diagram).

The results clearly indicate that the majority of examined countries in Africa are identified at a start-up stage of maturity. This translates into lack of a national programme for cybersecurity awareness raising. The need for awareness of cybersecurity threats and vulnerabilities across all sectors is not recognised, or is only at initial stages of discussion. Furthermore, awareness raising programmes (if existing) may be informed by international initiatives but are not linked to a national strategy.

Finally, it was identified that awareness raising programmes, courses, seminars and online resources might be available for target demographics from public, private, academic, and/or civil sources, but no coordination or scaling efforts have been conducted. In the next section, we provide further details, based on our qualitative analysis, on these initial findings.

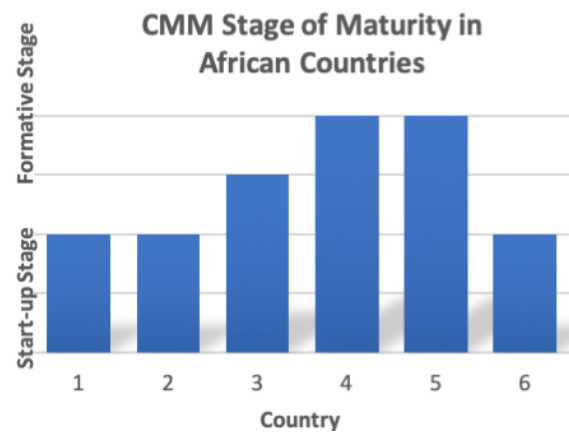

Figure 1. CMM results from six African Countries

### B. Qualitative analysis of results

We have transcribed all the recordings from focus groups and conducted a thematic analysis on the qualitative data for each country. We adopted a blended approach (a mix of deductive and inductive approach) to analyse focus group data and used the indicators of the CMM as our criteria for a deductive analysis. The inductive approach is based on 'open coding' meaning that the categories or themes are freely created by the researcher, while the deductive content analysis requires the prior existence of a theory to underpin the classification process.

Excerpts that did not fit into themes were further analysed to highlight additional issues that stakeholders might have raised during the focus groups or to inform our understanding on what the next steps should be for a country.

Overall, we identified eight themes in our qualitative analysis for every country. Four themes were based on the aspects described in the CMM model and four themes emerged from the inductive approach. The themes from the inductive approach pertained information on what actions African countries should implement next. Since these eight themes were common for all six countries, we merged the excerpts for each theme from every country. We further examined these excerpts to identify common areas which hindered progress in cybersecurity awareness raising as well as key actions which countries should implement next to improve their cybersecurity posture in awareness raising.

More specifically, the four main themes that emerged from the deductive approach are: a) the lack of national level programmes; b) the existence of ad-hoc initiatives; c) the relationship between ICT literacy (the ability to use digital technology and tools) and awareness and d) executive awareness. In a similar vein, the inductive approach identified four themes which revolved around the same concepts described in the deductive analysis; the difference being that excerpts in the inductive themes pertained information about recommendations and next steps.

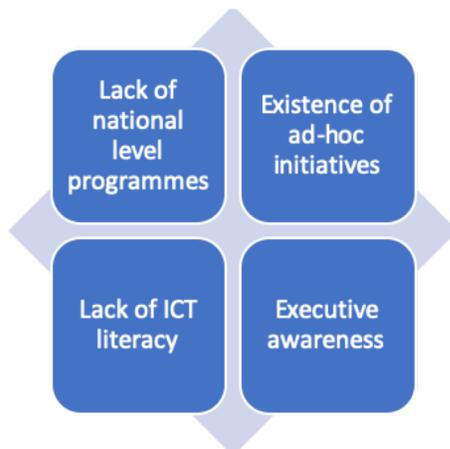

Figure 2. Themes from Inductive and Deductive Analysis

*1) Deductive theme analysis:* For all countries, it is evident that a national programme for cybersecurity awareness raising is absent. In many cases, stakeholders mentioned that '*lack of awareness is an institutional problem, not a user problem*' and also that '*a proper cyber awareness programme is needed*'. The importance for such a programme was acknowledged across the various stakeholders in all countries reviewed in Africa. A main hindrance for the implementation of a national programme is the general lack of cybersecurity awareness outside the technical communities, which stakeholders pointed that its origin is the low ICT literacy in the population of these countries.

Concerns were also expressed regarding the security of nationwide projects involving big volumes of personal data. Participants, mentioned that '*cybersecurity awareness, in particular in relation to the protection of personal data, needs to be prioritised for such projects*'.

It was further emphasised that awareness-raising programmes need to be developed alongside other capacity enhancements, such as incident response, training for cybersecurity educators, and national and organisational cybersecurity policies.

Regarding the initiatives theme, there are ad-hoc initiatives in cybersecurity awareness raising that are supported by various institutions. These are being offered from various organisations such as Facebook while the financial sector, civil society and academia organise programmes for schools to raise awareness. According to a stakeholder, '*some telecommunication companies and banks are engaged in awareness activities which includes messages via the media, directed to end-users, e.g. password security*'.

These initiatives, however, are not yet coordinated at the national level. Therefore, it was widely recognised that a more centralised awareness-raising programme would greatly expand a fundamental understanding of cybersecurity capacity.

Often, civil society actors initiate efforts into targeted cybersecurity awareness-raising. Different stakeholders agree that a '*common ground*' between government, private sector and civil society could enable the proliferation of awareness raising to the broader society. Moreover, often it was mentioned that the government needs to work alongside existing efforts in academia to ensure that new initiatives capitalise from the academic experience. Such synergy is critical to ensure that awareness-raising efforts are efficient and effective.

As often mentioned by stakeholders '*people trust social media and do not expect that someone will harm them, we are brothers!*'. A stakeholder also noted that '*It is common in African countries that mobile phones are used to access the Internet, use social media, for e-banking services etc. but people who use online services are not aware of risks*'. Often, lack of awareness leads to a sense of '*blind trust online*'. A stakeholder noted that '*users trust social media and think that their information is secure, although often websites are still insecure*'.

Another interesting theme that emerged from the analysis of data is the low ICT literacy rate in Africa. Stakeholders indicated that awareness of the effective use of ICT is still only gaining initial traction and that security is seen as only relevant once ICT and Internet literacy is sufficient. Stakeholders suggested that '*integrating cybersecurity awareness efforts into ICT literacy courses could provide an established vehicle for cybersecurity awareness-raising campaigns*'.

Regarding the theme revolving around awareness among executives, both in public and private sectors, cybersecurity awareness is very limited, which is one reason why cybersecurity awareness raising is not yet perceived as a priority. This has been identified as an important gap, as executives are usually the final arbiters on investment into security.

Some major telecommunications companies conduct internal awareness raising training across all levels, but there is not a publicly available initiative which targets executives. As mentioned by a stakeholder, '*the reason for that is that there is limited awareness for cybersecurity threats and risks in the private sector overall, unless in major international organi-sations, in particular in the banking and telecommunications sectors which face strategic implications of cybersecurity'.

It was commonly stated that there is a sharp disconnect between the terminology and priorities of the engineers working in IT systems and security, and those at the higher level seeking to make sound business decisions based on risk.

*2) Inductive Theme Analysis:* Stakeholders mentioned during focus group sessions that '*aspects of cybersecurity need to be introduced in the school curricula and improve ICT literacy*'. It was also noted that '*even in universities, people are not aware of the possible risks and procure without following standards*'. Integrating cybersecurity awareness efforts into ICT literacy courses could provide an established vehicle for cybersecurity awareness campaigns.

Culture is another factor that can impact the effectiveness of cybersecurity awareness programmes. As seen above, the collectivist cultural aspect that characterises off-line behaviour in Africa, is also pertained in online behaviour [44].

Currently, due to the lack of national level awareness programmes, '*being hacked brings awareness usually*' as a stakeholder noted. Therefore, the development of such a programme with specified target groups focusing on most vulnerable users is identified as necessary [45]. Also, appointing a designated organisation (from any sector) to lead the cybersecurity awareness raising programme and engaging relevant stakeholders from public and private sectors in the development and delivery of the awareness raising programme is crucial. As stakeholders mentioned in one of the reviews in Africa '*The government realises that lack of awareness is crucial and recognises the importance of a multi-stakeholder approach towards this goal*'. Moreover, it was noted that '*People access social media through their smart phones and security is the last thing on their mind and that convenience is usually coming first*'.

Stakeholders mentioned that '*even though the telecommunications sector has started to place emphasis on cybersecurity standards compliance, small- and medium-sized enterprises (SMEs) are mostly not worried about adopting and implementing standards*'. An area of particular concern for SMEs is that of encouraging good security behaviour by employees [46], [47], [48]. Developing a strong security culture could address many of the behavioural issues that underpin data breaches in such companies [49], [50]. Here, the development of cybersecurity skills involves addressing digital threats using technology and complementary factors including policy guidelines, organisational processes, and education and awareness strategies. By having an organisational security setting where employees intuitively protect corporate information assets, SMEs could improve their overall security [51].

Regarding the executive awareness raising aspect, developing a dedicated awareness raising programme for executives within the public and private sectors is essential. A stakeholder noted that '*different levels of authority need different kind of awareness in order to promote collaboration as well*'. Currently, executives and management are being called upon to address cyber risk alongside other risks that businesses face.

## V. DISCUSSION

Reflecting on the results presented in Section 4, the lack of a central authority, which is crucial in all modes of operation as presented by NIST model [16], is evident. The absence of such authority prohibits the execution of holistic 'needs assessments', amplifies the difficulties in prioritising the areas in which campaigns should be implemented and renders the design of ad-hoc campaigns being created by a limited number of stakeholders. It is imperative that African countries allocate an authority to conduct a national needs assessment, identify the areas where campaigns should focus first, develop a strategy for how these campaigns will be designed and implemented, and coordinate the ad-hoc efforts of different stakeholders.

The main objectives for cybersecurity in Africa and globally is online security by improving knowledge, capabilities and decision making. In order to enable the full benefits of cyberspace to all African countries, investing in human capacity development of all the citizens is vital.

Focusing on the design and implementation of awareness-raising campaigns, literature suggests that successful awareness campaigns need to be a 'learning continuum' [52], commencing from awareness, evolving to training and resulting in education. Our results highlight the need of African countries to involve stakeholders who are established in all the aforementioned sectors. Our analysis suggests that the audience of the campaigns should prioritise smartphone users, employees of SMEs and board members. The goals should be to communicate the risks from cybercrime, illustrate the need for better security controls and practices, and the need to establish a chief information security officer (CISO), respectively.

This means that businesses and government agencies should start to take steps to increase their awareness and understanding of cybersecurity with a view of the potential impact on overall business performance. Lack of boardroom expertise makes it challenging for directors and councilors to effectively oversee management's cybersecurity activities.

Cybersecurity awareness should reach all levels and inform all users of the internet - from vulnerable, school-going children to families, industry, critical national infrastructures, governments and the African continent with its unique needs [53], [54], [55]. This will enhance resilience against cybercrimes and attacks and inform African policy development.

If a country has already developed a national cybersecurity strategy, or is working towards that goal, then linking the development of the programme to that Strategy will facilitate the coordination of different capacities towards the development of the programme and its effective implementation.

Regarding the implementation of these campaigns, there are several organisations with ad-hoc initiatives that could facilitate the design and implementation of cybersecurity campaigns, such as ISC Africa [28] and Parents corner [29]. To conclude, it is worth mentioning that the timing for the development of these campaigns coincides with efforts in African countries to increase ICT literacy. As our findings underline, it is a unique opportunity for all African countries to combine ICT development with cybersecurity awareness. In contrast to western societies, where cybersecurity campaigns endeavour to change the norms on how users currently behave online (behaviour shaped since the inception of the Internet), campaigns in Africa can reflect on best practice and create new norms which will encompass cybersecurity requirements.

Creating a single online portal linking to appropriate cybersecurity information and disseminating it via the cybersecurity awareness programme can also enhance the effectiveness of such a programme. Moreover, enacting evaluation measurements to study the effectiveness of an awareness programme will not only lead to the assessment of the programme but also identify possible gaps that need to be addressed [45].

Moreover, enacting evaluation measurements to study effectiveness of the awareness programme will not only lead to the assessment of the programme but also identify possible gaps that need to be addressed [16], [45].

## VI. Conclusions and future work

Several reports are depicting a bleak picture regarding the unprecedented increase of cybercrime in Africa. Yet, efforts to raise cybersecurity awareness in the general public and executives are in an embryonic stage. In this paper, we conducted twelve focus groups in six different African countries to shed light into the current situation and identify critical actions which can significantly decrease the success rate of cybercriminals.

Our results suggest that all six African countries do not possess a national programme for raising awareness, there are extremely low ICT literacy levels which hinder any design of cybersecurity campaigns and that executive members in organisations myopically underestimate the problem. To better defend against cybercrime, African countries need to establish a central authority which will coordinate the existing ad-hoc efforts in awareness campaigns and identify the target groups of these campaigns with particular focus on SMEs, mobile-phone users and executive board members. We believe that African countries have a unique opportunity to combine ICT literacy campaigns with cybersecurity principals and shape the norms of the society towards best practice.

By improving knowledge, cybersecurity can also be enhanced as well as capabilities and decision making. In Africa, but also at the global level the full benefits of cyberspace can be enabled by investing in human capacity development. Executives are also users and they need also to be aware of how cyber risks can threaten their assets in order to make effective strategy decisions.

At a national and an organisational level, strategies need to be developed linked to awareness campaigns with clear objectives, design and implementation processes and coordination of the ad-hoc efforts of different stakeholders. As part of our future work, we intend to explore the effectiveness of a national coordinated cybersecurity awareness programme and how it relates to the actual security posture of a country. Our future work will be based on data from developed countries where the CMM has already been applied, as well as on data collected by other international organisations such as the International Telecommunication Union - GCI [56], Australian Strategic Policy Institute - ASPI [57], The Potomac Institute for Policy Studies - Cyber Readiness Index [58] and World Economic Forum - Global Competitive Index [59].

## *Acknowledgment

The authors would like to thank Ms. Eva Ignatuschtschenko, Ms. Eva Nagyfejeo, Mr. Taylor Roberts and Ms. Carolin Weisser from the GCSCC for conducting field work and data collection. We are also immensely grateful to Prof. Sadie Creese and Prof. Michael Goldsmith for their comments on an earlier version of the manuscript.

APPENDIX

In this section we present the details of the capacity maturity model for dimension 3 used to analyse the results of the qualitative research.

| Aspect | Start-Up | Formative | Established | Strategic | Dynamic |
|---|---|---|---|---|---|
| | **D 3.1: Awareness Raising** | | | | |
| Awareness Raising Programmes | The need for awareness of cybersecurity threats and vulnerabilities across all sectors is not recognised, or is only at initial stages of discussion. | Awareness raising programmes, courses, seminars and online resources are available for target demographics from public, private, academic, and/or civil society sources, but no coordination or scaling efforts have been conducted. Awareness raising programmes may be informed by international initiatives but are not linked to national strategy. | A national programme for cybersecurity awareness raising, led by a designated organisation (from any sector) is established, which addresses a wide range of demographics and issues, but no metrics for effectiveness have been applied. Consultation with stakeholders from various sectors informs the creation and utilisation of programmes and materials. A single online portal linking to appropriate cybersecurity information exists and is disseminated via that programme. | The national awareness raising programme is coordinated and integrated with sector-specific, tailored awareness raising programmes, such as those focusing on government, industry, academia, civil society, and/or children. Metrics for effectiveness are established and evidence of application and lessons learnt are fed into future programmes. The evolution of the programme is supported by the adaptation of existing materials and resources, involving clear methods for obtaining a measure of suitability and quality. Programmes contribute toward expanding and enhancing international awareness raising good practice and capacity-building efforts. | Awareness raising programmes are adapted in response to performance evidenced by monitoring which results in the redistribution of resources and future investments. Metrics contribute toward national cybersecurity strategy revision processes. Awareness programme planning gives explicit consideration to national demand from the stakeholder communication (in the widest sense), so that campaigns continue to impact the entire society. The national awareness raising programme has a measurable impact on reduction of the overall threat landscape. |

Figure 3. Dimension 3: Cybersecurity Education, Training and Skills for Awareness Raising Programmes

| Aspect | D 3.1: Awareness Raising | | | | |
|---|---|---|---|---|---|
| | Start-Up | Formative | Established | Strategic | Dynamic |
| Executive Awareness Raising | Awareness raising on cybersecurity issues for executives is limited or non-existent. Executives are not yet aware of their responsibilities to shareholders, clients, customers, and employees in relation to cybersecurity. | Executives are made aware of general cybersecurity issues, but not how these issues and threats might affect their organisation. Executives of particular sectors, such as finance and telecommunications, have been made aware of cybersecurity risk in general and how the organisation deals with cybersecurity issues, but not of strategic implications. | Awareness raising of executives in the public, private, academic and civil society sectors address cybersecurity risks in general, some of the primary methods of attack, and how the organisation deals with cyber issues (usually abdicated to the CIO). Select executive members are made aware of how cybersecurity risks affect the strategic decision making of the organisation, particularly those in the financial and telecommunications sectors. Awareness raising efforts of cybersecurity crisis management at the executive level is still reactive in focus. | Executive awareness raising efforts in nearly all sectors include the identification of strategic assets, specific measures in place to protect them, and the mechanism by which they are protected. Executives are able to alter strategic decision making, and allocate specific funding and people to the various elements of cyber risk, contingent on their company's prevailing situation. Executives are made aware of what contingency plans are in place to address various cyber-based attacks and their aftermath. Executive awareness courses in cybersecurity are mandatory for nearly all sectors. | Cybersecurity risks are considered as an agenda item at every executive meeting, and funding and attention is reallocated to address those risks. Executives are regarded regionally and internationally as a source of good practice in responsible and accountable corporate cybersecurity governance. |

Figure 4. Dimension 3: Cybersecurity Education, Training and Skills for Executive Awareness Raising